\documentstyle[NATO,psfig]{crckapb}
\begin{opening}
\title{Superconductivity in Ultrasmall Grains: 
  \protect\\ Introduction to Richardson's Exact Solution}
\author{Jan von Delft and Fabian Braun}
\institute{Institut f\"ur Theoretische Festk\"orperphysik \\
Universit\"at Karlsruhe \\ 76128 Karlsruhe, Germany\\
{\rm November 1, 1999}}
\runningtitle{Richardson's Exact Solution of a Reduced BCS Model}
\end{opening}
\begin{document}
  \begin{abstract}
    Studies of pairing correlations in ultrasmall metallic grains have
    commonly been based on a simple reduced BCS-model describing the
    scattering of pairs of electrons between discrete energy levels
    that come in time-reversed pairs. This model has an exact
    solution, worked out by Richardson in the context of nuclear
    physics in the 1960s.  Here we give a tutorial introduction to his
    solution, and use it to check the quality of various previous
    treatments of this model.
 \end{abstract}

\section{Introduction}

Recent experiments by Ralph, Black and Tinkham, involving the
observation of a spectroscopic gap indicative of pairing correlations
in ultrasmall Al grains \cite{RBT}, have inspired a number of
theoretical \cite{vD}-
\cite{DP} studies of
how superconducting pairing correlations in such grains are affected
by reducing the grains' size, or equivalently by increasing its mean
level spacing $d \propto {\rm Vol}^{-1}$ until it exceeds the bulk gap
$\Delta$.  In the earliest of these, a grand-canonical (g.c.)  BCS
approach \cite{vD,Braun1,BvD} was applied to a reduced BCS Hamiltonian
for uniformly spaced, spin-degenerate levels; it suggested that
pairing correlations, as measured by the condensation energy $E^C$,
vanish abruptly once $d$ exceeds a critical level spacing $d^c$ that
depends on the parity (0 or 1) of the number of electrons on the
grain, being smaller for odd grains ($d^c_1 \simeq 0.89 \Delta$) than
even grains $(d^c_0 \simeq 3.6 \Delta$).  A series of more
sophisticated canonical approaches (summarized in
Section~\ref{comparison} below) confirmed the parity dependence of
pairing correlations, but established 
\cite{ML}-
\cite{DP}
that the abrupt vanishing of pairing correlations at $d^c$ is an
artifact of g.c.\ treatments: pairing correlations do persist, in the
form of so-called fluctuations, to arbitrarily large level spacings,
and the crossover between the bulk superconducting (SC) regime $(d \ll
\Delta)$ and the fluctuation-dominated (FD) regime $(d \gg \Delta)$ is
completely smooth \cite{DS}. Nevertheless, these two regimes are
qualitatively very different \cite{Braun2,DS}: the condensation
energy, e.g., is an extensive function of volume in the former and
almost intensive in the latter, and pairing correlations are quite
strongly localized around the Fermi energy $\varepsilon_F$, or more
spread out in energy, respectively.

After the appearance of all these works, we became aware that the
reduced BCS Hamiltonian on which they are based actually has an exact
solution. It was published by R. W. Richardson in the context of nuclear
physics (where it is known as the ``picket-fence model''), in a series
of papers between 1963 and 1977
\cite{richardson63a}-
\cite{richardson77} which until
very recently seem to have completely escaped the attention of the
condensed matter community.  In this work, we
(i) give a tutorial
   introduction  (with no pretense of rigor) to his solution,
and (ii) compare the results of various previously-used approximations
against the benchmark set by the exact solution, in order to gauge
their reliability for related problems for which no exact solutions
exist \cite{Braun-thesis,sierra-vd}.

\section{Richardson's Exact Solution}

\subsection{Reduced BCS Model}
Ultrasmall superconducting grains 
are commonly described \cite{vD}-
\cite{DP}   by a reduced BCS model,
\begin{equation}
  \label{eq:H1}
  H = \sum_{j\sigma} \varepsilon_j c^\dagger_{j\sigma} c_{j \sigma}
    - g \sum_{ij}  c^\dagger_{i+}  c^\dagger_{i-} c_{j-} c_{j+} ,
\end{equation}
for a set $S$ of $N_S$ pairs of time-reversed states $|j, \pm \rangle$
labeled by a discrete index $j= 1, \dots, N_S$, with energies
$\varepsilon_j$ and coupling $g =\lambda \, d$, where $d$ is the mean
level spacing and $\lambda $ a dimensionless coupling constant.
Unbeknownst to the authors that have studied this model recently,
Richardson had long ago solved it exactly, for an arbitrary
set of levels $\varepsilon_j$ (degenerate levels are allowed, but are
to be distinguished by distinct $j$-labels, i.e.\ they have
$\varepsilon_i = \varepsilon_{j}$ for $i \neq j$). 

The first step is to note that {\em singly-occupied}\/ levels do not
participate in the pairscattering described by $H$, and by the Pauli
principle remain ``blocked'' \cite{Soloviev} to such pairscattering;
the labels of such levels are therefore good quantum numbers. A
general eigenstate of $H$ thus has the form
\begin{eqnarray}
|n,B\rangle & = &  \prod_{i \in B} c_{i \sigma}^\dagger |\Psi_n\rangle_U ,
  \label{eq:generaleigenstate} 
\\
  \label{eq:generaleigenstate-2} 
 |\Psi_n \rangle_U & = & \sum_{j_1, \dots, j_n}^U
\psi (j_1, \dots , j_n) \prod_{\nu=1}^n b_{j_\nu=1}^\dagger  |0 \rangle
\; .
\end{eqnarray}
This describes $N = 2n + b$ electrons, $b$ of which sit in a set $B$ of
singly-occupied, blocked levels, thereby contributing ${\cal E}_B = 
\sum_{i \in B} \varepsilon_i$ to the eigenenergy,
while the remaining $n$ pairs of electrons, created by the pair
operators $b_j^\dagger = c^\dagger_{j +} c^\dagger_{j -}$, are
distributed among the remaining set $U= S \backslash B$ of $N_U=
N_S-b$ {\em unblocked} levels, with wave function $\psi (j_1, \dots ,
j_n)$ ($\sum_j^U \equiv \sum_{j \not \in B}$ denotes a sum over
all {\em unblocked\/} levels).  The dynamics of these pairs is
governed by
\begin{equation}
 H_U = \sum_{ij}^U \left( 2 \varepsilon_j \delta_{ij} -
 \; g \right)  b_i^\dagger b_j \; ,
 \label{1}
\end{equation}
and writing the eigenenergy of $|n,b\rangle$ as ${\cal E}_n + {\cal E}_b$,
the state $|\Psi_n\rangle_U$ satisfies
\begin{eqnarray}
  \label{eq:eigenpsi}
   H_U |\Psi_n\rangle_U = {\cal E}_n |\Psi_n\rangle_U , \qquad
   \sum_j^U b_j^\dagger b_j |\Psi_n\rangle_U = n |\Psi_n\rangle_U \; .
\end{eqnarray}
Diagonalizing $H_U$ would be trivial if the $b$'s were true bosons.
However, they are not, and in the subspace spanned by the set $U$ of
all non-singly-occupied levels, instead satisfy the ``hard-core
boson'' relations,
\begin{eqnarray}
\label{hard-core-boson-1}
  b^{\dagger 2}_j = 0, \qquad
\label{hard-core-boson-2}
\mbox{[} b_j, b^\dagger_{j^{\prime}}\mbox{]} =
\delta_{j j^{\prime}} (1 - 2 b^\dagger_j b_j), \qquad 
\label{hard-core-boson-3}
\mbox{[} b^\dagger_{j} b_{j},  
b^\dagger_{j'} \mbox{]} &=& \delta_{j j'} b^\dagger_j \; ,
\end{eqnarray}
which reflect the Pauli principle for the fermions they are
constructed from. In particular, $b^{\dagger 2}_j=0$ implies that only
those terms in (\ref{eq:generaleigenstate-2}) are non-zero for which
the indices $j_1, \dots j_n$ are all distinct.

In his original publications
\cite{richardson63a,richardson63b,richardson64}, Richardson derived a
Schr\"odinger for $\psi (j_1, \dots , j_n)$ and showed that its exact
solution was simply a generalization of the form that $\psi (j_1,
\dots , j_n)$ would have had if the $b$'s had been true (not
hard-core) bosons.  With the benefit of hindsight, we shall here
follow an alternative, somewhat shorter root, also due to
Richardson\cite{richardson-priv}: we first consider the related but
much simpler case of {\em true\/} bosons and write down the generic form of
its eigenstates; we then clarify why this form fails to produce
eigenstates of the {\em hard-core\/} boson Hamiltonian; and
having identified the reason for the failure, we show that (remarkably)
only a slight generalization is needed to repair it and to obtain the
sought-after hard-core-boson eigenstates.

\subsection{True Bosons}
Let $\tilde b_j$ denote a set of true bosons (i.e.\
$[ \tilde b_j, \tilde b_{j'}^\dagger] = \delta_{j j'}$), 
governed by a Hamiltonian $\tilde H_U$ of precisely the form
(\ref{1}), with  $b_j \to \tilde b_j$. 
 This problem, being quadratic, can be solved straightforwardly by
any  number of methods. The solution is as follows: 
$\tilde H_U$ can be written as
\begin{eqnarray}
  \label{eq:trueosons}
  \tilde H_U = \sum_J \tilde E_J \tilde B_J^\dagger \tilde B_J  + \mbox{const.}
\end{eqnarray}
where the new bosons $\tilde B_J^\dagger$ 
(with normalization constants $C_J$) are given by 
\begin{eqnarray}
  \label{eq:newbosons}
 \tilde B^\dagger_J 
= g C_J \sum_j^U {\tilde b^\dagger_j \over 2 \varepsilon_j -  \tilde E_J}
 \; , \qquad
  \label{eq:normalization}
  {1 \over (g C_J)^2} = 
\sum_j^U {1 \over (2 \varepsilon_j -  \tilde E_J)^2} \; ,
\end{eqnarray} 
and the boson eigenenergies $ \tilde E_J$ are the roots of the eigenvalue
equation
\begin{eqnarray}
  \label{eq:eigenvalue}
  1 - \sum_j^U {g \over 2 \varepsilon_j -  \tilde E_J} \; = \; 0 \; .
\end{eqnarray}
This is an equation of order $N_U$ in $ \tilde E_J$. It thus has $N_U$
roots, so that the label $J$ runs from 1 to $N_U$.  As the coupling
$g$ is turned to 0, each $E_J$ smoothly evolves to one of the bare
eigenenergies $\varepsilon_j$.  A general $n$-boson eigenstate of $\tilde H_U$
and its eigenenergy $\tilde {\cal E}_n$ thus have the form
\begin{eqnarray}
  \label{eq:truebosoneigenstates-1}
  |\tilde \Psi_n \rangle_U = \prod_{\nu=1}^n \tilde 
B_{J_\nu}^\dagger |0\rangle \, ,
\qquad
\label{Etotal} 
\tilde {\cal E}_n = \sum_{\nu=1}^n  \tilde E_{J_\nu} \; ,
\end{eqnarray} 
where the $n$ indices $J_1, \dots, J_n$ that characterize this state need
not all be distinct, since the $B_J^\dagger$ are true bosons.

\subsection{Complications Arising for Hard-Core Bosons}
Let us now return to the hard-core boson Hamiltonian $H_U$.  Its
eigenstates will obviously {\em not\/} be identical to the true-boson
eigenstates just discussed, since matters are changed considerably by
the hard-core properties of $b_j$.  To find out exactly {\em what\/}
changes they produce, it is very instructive to take an Ansatz for $|
\Psi_n \rangle_U$ similar to (\ref{eq:truebosoneigenstates-1}) (but
suppressing the normalization constants and taking all $J_{\nu}$ to be
distinct), namely
\begin{eqnarray}
  \label{eq:truebosoneigenstates}
  | \Psi_n \rangle_U = \prod_{\nu=1}^n 
B_{J_\nu}^\dagger |0\rangle \, ,
\qquad \mbox{with} \qquad 
  B_J^\dagger = \sum_j^U {b_j^\dagger \over 2 \varepsilon_j - E_J} \; ,
\end{eqnarray} 
and to check explicitly whether or not it could be an eigenstate of $H_U$,
i.e.\ to check under what conditions $(H_U - {\cal E}_n) |\Psi_n \rangle_U $
would equal zero, where ${\cal E}_n = \sum_{\nu}^n E_{J_\nu}$.  To this end,
we commute $H_U$ to the right past all the $B^\dagger_{J_\nu}$ operators in
$|\Psi_n \rangle_U$, using
\begin{eqnarray}
  \label{eq:commuteHtofront-1}
 \left[ H_U,  \prod_{\nu = 1}^{n} B_{J_\nu}^\dagger  \right] =
\sum_{\nu = 1}^n \left\{ 
\left( \prod_{\eta = 1}^{\nu -1} B_{J_\eta}^\dagger \right)
[H_U,  B_{J_\nu}^\dagger ] 
\left( \prod_{\mu = \nu+1}^n B_{J_\mu}^\dagger \right) \right\} .
\end{eqnarray}
To evaluate the commutators appearing here, we write
$H_U$ as 
\begin{eqnarray}
  \label{eq:checkAnsatztrueboson}
  H_U = \sum_{j}^U 2 \varepsilon_j  b_j^\dagger b_j \; - \;
   g B^\dagger_0  B_0 \; , \qquad
 \mbox{where} \quad B^\dagger_0 = \sum_j^U  b^\dagger_j \; , 
\end{eqnarray}
and use the following relations:
\begin{eqnarray}
  \label{eq:bjBJ}
  [b_j^\dagger b_j, B^\dagger_J] = 
     {b_j^\dagger \over    2 \varepsilon_j - E_J} \; , \qquad
  [B_0, B^\dagger_J] = \sum_j^U
     {1- 2 b_j^\dagger b_j \over    2 \varepsilon_j - E_J} \; , 
\end{eqnarray}
\begin{eqnarray}
\label{HUBJ}
  [H_U, B^\dagger_J] =  E_J B_J^\dagger \; + \;
     B_0^\dagger \left[ 1 - g \sum_j^U {1 - 2 b_j^\dagger b_j 
  \over 2 \varepsilon_j - E_J}    \right] \; .
\end{eqnarray}
Inserting these into (\ref{eq:commuteHtofront-1}) and using $H_U
|0\rangle =0$ and ${\cal E}_n = \sum_{\nu}^n E_{J_\nu}$, we find
\begin{eqnarray}
\nonumber
\lefteqn{\phantom{.} \hspace{-1.5cm}
  H_U |\Psi_n\rangle_U  =   {\cal E}_n |\Psi_n\rangle_U 
+ \sum_{\nu=1}^n \left[ 1 - \sum_j^U 
{g  \over 2 \varepsilon_j - E_{J_\nu}}    \right]
\! B_0^\dagger \! 
\left( \prod_{\eta =1 (\neq \nu)}^{n} B_{J_\eta}^\dagger \right) |0\rangle} 
\qquad \qquad \qquad \phantom{.} \\
  \label{eq:commuteHtofront-2}
& & \phantom{.} \hspace{-3.5cm} + \sum_{\nu=1}^n
\left\{ \left( \prod_{\eta = 1}^{\nu -1} B_{J_\eta}^\dagger \right)
\left[   \sum_j^U { 2 g B_0^\dagger  \, b_j^\dagger b_j 
  \over 2 \varepsilon_j - E_{J_\nu} }    \right] 
\left( \prod_{\mu = \nu+1}^n B_{J_\mu}^\dagger \right) \right\} |0\rangle \; .
\end{eqnarray}
Now, suppose we do the same calculation for true instead of hard-core
bosons (i.e.\ run through the same steps, but place a $\tilde
{\phantom{.}}$ on $H_U$, $b_j$,  $E_J$ and ${\cal E}_n$). 
Then the second line of
(\ref{eq:commuteHtofront-2}) would be absent (because the
$b_j^\dagger b_j$ terms in the second of
Eqs.~(\ref{hard-core-boson-2}) and (\ref{eq:bjBJ})
and in (\ref{HUBJ}) would be absent); and 
the first line of (\ref{eq:commuteHtofront-2}) would imply that 
$(\tilde H_U - \tilde {\cal   E}_n) |\tilde \Psi_n\rangle_U = 0$
provided that the term in square brackets vanishes, 
which is nothing but the condition that  the $\tilde E_J$
satisfy the 
the true-boson eigenvalue equation of
(\ref{eq:eigenvalue})! In other words, we have just verified
explicitly that all true-boson states of the form
(\ref{eq:truebosoneigenstates-1}) are indeed eigenstates of $\tilde
H_U$, provided that the $\tilde E_J$ satisfy (\ref{eq:eigenvalue}).
Moreover, we have identified the term in second line of
(\ref{eq:commuteHtofront-2}) as the extra complication that arises for
hard-core bosons.

\subsection{The Cure: a Generalized Eigenvalue Equation}

Fortunately, this extra complication is tractable:
first, we note that 
\begin{eqnarray}
  \label{eq:newtermBJ}
  \left[   \sum_j^U { 2 g B_0^\dagger \, b_j^\dagger b_j 
  \over 2 \varepsilon_j - E_{J_\nu}} , B^\dagger_{J_\mu}   \right]
 =  \sum_j^U {  2 g B_0^\dagger 
  \over 2 \varepsilon_j - E_{J_\nu}} 
 {  b_j^\dagger \over 2 \varepsilon_j - E_{J_\mu}}
= 2  g B_0^\dagger \, {B_{J_\nu}^\dagger - B_{J_\mu}^\dagger
  \over  E_{J_\nu} -  E_{J_\mu}}  .
\end{eqnarray}
The rightmost expression follows via a partial fraction expansion,
and remarkably, contains only $B^\dagger$ operators
and no more $b^\dagger_j b_j$s. This enables
us to eliminate the  $b^\dagger_j b_j$s from 
the second line of (\ref{eq:commuteHtofront-2}), by
rewriting it  as follows (we commute its
term in square brackets to the right, using a
relation similar to (\ref{eq:commuteHtofront-1}),
but with the commutator (\ref{eq:newtermBJ}) instead
of $[H_U,  B_{J_\mu}^\dagger]$):
\begin{eqnarray}
&& \phantom{.} \hspace*{-1cm} 
\sum_{\nu=1}^n \!
\left\{ \! \! \left( \prod_{\eta = 1}^{\nu-1} B_{J_\eta}^\dagger \right)\!\!
\sum_{\mu = \nu +1}^n \! \!
\left\{ \! \! 
\left( \prod_{\eta' = \nu + 1}^{\mu -1} \!\! B_{J_{\eta'}}^\dagger \right) \!\!
\left[   2  g B_0^\dagger \, {B_{J_\nu}^\dagger - B_{J_\mu}^\dagger
  \over  E_{J_\nu} -  E_{J_\mu}} \right] \!\!
\left( \prod_{\mu' = \mu +1}^n \!\! B_{J_{\mu'}}^\dagger
\right) \! \! \right\} \! \! \right\}  \!
|0\rangle \nonumber \\
&& = 
\sum_{\mu = 1}^n \! \!
\left[ \sum_{\nu = 1}^{\mu - 1} {2g \over E_{J_\nu} - E_{J_{\mu}}} \right]
\! B_0^\dagger \! \left( \prod_{\eta =1 (\neq  \mu)}^{n} 
\!\! B_{J_\eta}^\dagger \right) |0 \rangle \nonumber
\\
&& \qquad  - \sum_{\nu = 1}^n \! \!
\left[ \sum_{\mu = \nu+1}^n {2g \over E_{J_\nu} - E_{J_{\mu}}} \right]
\! B_0^\dagger \! \left( \prod_{\eta =1 (\neq  \nu)}^{n} 
\!\! B_{J_\eta}^\dagger \right)  |0 \rangle   \nonumber \\
&&= 
   \sum_{\nu = 1}^n \! \!
\left[ \sum_{\mu =1 (\neq \nu)}^n {2g \over E_{J_\mu} - E_{J_{\nu}}} \right]
B_0^\dagger \left( \prod_{\eta =1 (\neq  \nu)}^{n} 
\!\! B_{J_\mu}^\dagger \right) |0 \rangle  \; . 
  \label{eq:commuteHtofront-3}
\end{eqnarray}
(The last line follows by renaming the dummy indices
$\nu \leftrightarrow \mu$ in the second line.)
Substituting (\ref{eq:commuteHtofront-3}) for the second line of
(\ref{eq:commuteHtofront-2}),  we conclude that $ ( H_U - {\cal E}_n)
|\Psi_n\rangle_U$ will equal zero provided that
\begin{eqnarray}
  \label{eq:richardson-eigenvalues}
  1 - \sum_j^U 
{g  \over 2 \varepsilon_j - E_{J_\nu}}    
+ \sum_{\mu =1 (\neq \nu)}^n {2g \over E_{J_{\mu}} - E_{J_{\nu}}} = 0 \; ,
\qquad \mbox{for}\quad \nu = 1, \dots, n \; .
\end{eqnarray}
This consitutes a set of $n$ coupled equations for the $n$ parameters
$E_{J_1}, \dots,$ $E_{J_n}$, which may be thought of as
self-consistently-determined pair energies.
Eq.~(\ref{eq:richardson-eigenvalues}) can be regarded as a
generalization of the true-boson eigenvalue equation
(\ref{eq:eigenvalue}), and was originally derived by Richardson by
solving the Schr\"odinger equation for the wave-function $\psi (j_1,
\dots , j_n)$ of (\ref{eq:generaleigenstate-2}). It is truly
remarkable that the exact eigenstates of a complicated many-body
problem can be constructed by such a simple generalization of the
solution of a quadratic (i.e.\ non-interacting) true-boson
Hamiltonian!

Below we shall always assume the $\varepsilon_j$'s to be all distinct.
Then there exists a simple relation between the bare pair energies
$2\varepsilon_j$ and the solutions of
(\ref{eq:richardson-eigenvalues}): as $g$ is reduced to 0, it follows
by inspection that each solution $\{ E_{J_1}, \dots,$ $E_{J_n}\}$
reduces smoothly to a certain set of $n$ bare pair energies, say $\{2
\varepsilon_{j_1}, \dots, 2\varepsilon_{j_n} \}$.  Correspondingly,
the state $|\Psi_n \rangle_U \equiv |J_1,\dots J_n\rangle_U$ of
(\ref{eq:truebosoneigenstates}) reduces smoothly to the state $|j_1,
\dots j_n \rangle_U \equiv \prod_{\nu=1}^n b_{j_\nu}^\dagger |0
\rangle $ (up to a normalization factor not shown here).  Thus there
is a one-to-one correspondence between the set of all states $\{ |J_1,
\dots, J_n \rangle_U \} $ and the set of all states $\{ | j_1, \dots
j_n \rangle_U \}$. Since the latter constitute a complete eigenbasis
for the $n$-pair Hilbert space defined on the set of unblocked levels
$U$, the former do too.

\subsection{Ground State}

For a given set of blocked levels $B$, the lowest-lying of all states
$|n,B\rangle$, say $|n,B\rangle_{G}$, is obtained by using that particular
solution $E_{J_1}, \dots E_{J_n}$ for which the total ``pair energy'' ${\cal
  E}_n$ takes its lowest possible value (as $g$ is increased, some of the
$E_J$s become complex; however, they always occur in complex conjugate pairs,
so that ${\cal E}_n$ remains real \cite{richardson66}).

The lowest-lying of all eigenstates with $n$ pairs and $b$ blocked
levels, say $|n,b \rangle_G$ with energy ${\cal E}^G_b (n)$, is that
$|n, B \rangle_G$ for which the blocked levels in $B$ are all as close
as possible to $\varepsilon_F$, the Fermi energy of the uncorrelated
$N$-electron Fermi sea $|F_N\rangle$.  The $E_{J_\nu}$ for the
ground state $|n,b\rangle_G$ coincide at $g=0$ with the lowest $n$
energies $2 \varepsilon_j$ ($j = 1, \dots, n$), and smoothly evolve
toward lower values as $g$ is turned on.  This fact can be exploited
during the numerical solution of (\ref{eq:richardson-eigenvalues}),
which can be simplified by first making some algebraic
transformations, discussed in detail in \cite{richardson65a}, that
render the equations less singular.

\subsection{General Comments}
Since the exact solution provides us with wave functions, it is in principle
straightforward to calculate arbitrary correlation functions. Some such
correlators are discussed by Richardson in \cite{richardson65b,richardson66},
who showed that they can be expressed in terms of certain determinants that
are most conveniently calculated numerically.  Moreover, it is natural to ask
whether in the bulk limit, the standard BCS results can be extracted from the
exact solution.  Indeed they can, as Richardson showed in \cite{richardson77},
by interpreting the problem of solving (\ref{eq:richardson-eigenvalues}) for
the $E_{J_\nu}$ as a problem in two-dimensional electrostatics.
%
%
Exploiting this analogy, he showed that in the bulk limit ($N_S \to
\infty$ at at fixed $N_S d$), Eqs.~(\ref{eq:richardson-eigenvalues})
reduce to the well-known BCS gap equation and the BCS equation for the
chemical potential, and the condensation energy 
${\cal E}^C_0(n)$ (defined in Eq.~(\ref{eq:condensation}) below) 
to its BCS result,  namely $- \Delta^2/2d$.

\section{Comparison with Other Approaches}
\label{comparison}
We now apply the exact solution to check the quality of results
previously obtained by various other methods. Most previous works
\cite{vD,Braun1,BvD,ML,MFF,BH,Braun2,DS} studied a
half-filled band 
with fixed width $ 2\omega_D$ of uniformly-spaced levels (i.e.\ $\varepsilon_j
= j \, d$), containing $N=2n+b$ electrons.  Then the level spacing is $d= 2
\omega_D /N$ and in the limit $d \to 0$ the bulk gap is $\Delta = \omega_D
\sinh (1/\lambda)^{-1}$. Following \cite{Braun2}, we take $\lambda = 0.224$
throughout this paper.  To study the SC/FD crossover, two types of quantities
were typically calculated as functions of increasing 
$d/\Delta$, which mimics decreasing
grain size:  the even and odd ($b=0,1$) condensation energies
\begin{equation}
  \label{eq:condensation}
E^C_b (n) = {\cal E}^G_b (n) -
\langle F_N| H |F_N\rangle \; ;
\end{equation}%
\begin{figure}
\centerline{\psfig{figure=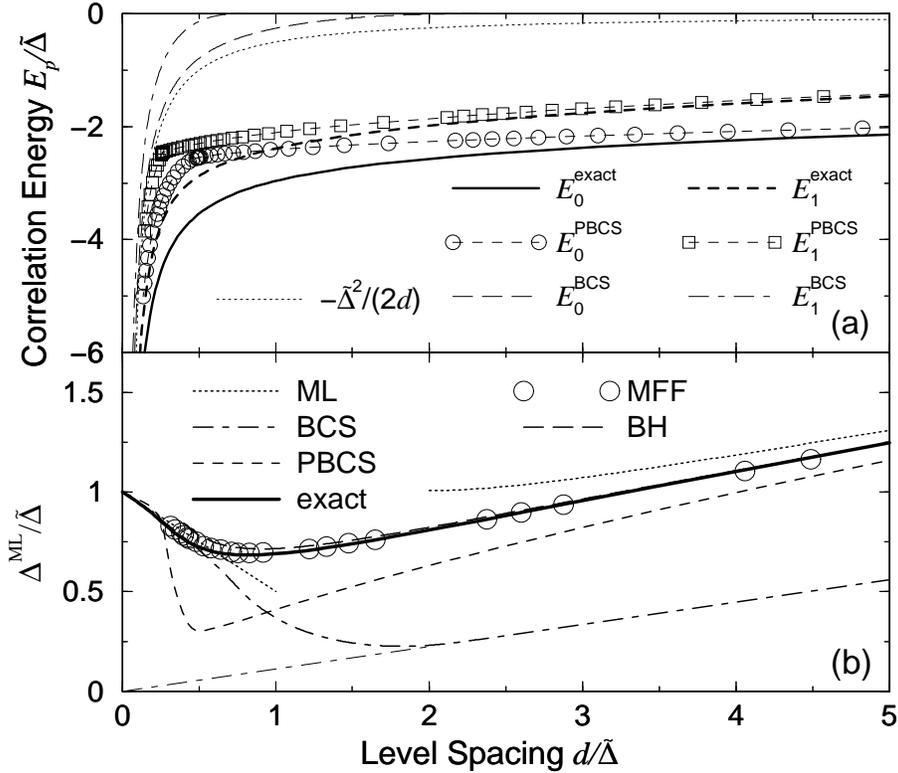,width=0.95\linewidth}} 
  \caption[Exact ground state condensation energies]{
    (a) The even and odd  $(b=0,1)$ condensation energies ${E}^C_b$ of
    Eq.~(\ref{eq:condensation}),
   calculated with BCS, PBCS and exact wave
    functions, as functions of $d/  \Delta = 2 \sinh (1/ \lambda) / (2n
    + b)$, for $\lambda = 0.224$.  For comparison the dotted line gives the
    ``bulk'' result $E_0^{\rm bulk} = -\Delta^2/(2d)$.  (b) Comparison
    of the parity parameters $\Delta^{\rm ML}  $ \protect\cite{ML} 
   of Eq.~(\ref{eq:ML}) obtained
    by various authors: ML's analytical result (dotted lines)
    [$\Delta(1-d/2 \Delta)$ for $d\ll \Delta$, and $d/2\log(a
    d/\Delta)$ for $d\gg \Delta$, with $a=1.35$ adjusted to give
    asymptotic agreement with the exact result]; grand-canonical
    BCS approach (dash-dotted line) [the naive perturbative result
    $\frac12\lambda d$ is continued to the origin]; PBCS approach
    (short-dashed line); Richardson's exact solution (thick solid line); exact
    diagonalization and scaling by MFF (open circles) and BH (long-dashed
    line).  }
  \label{fig:exact-gse}
\end{figure}%
\phantom{.} \hspace{-6.5mm} 
and a parity parameter introduced by Matveev and Larkin (ML) \cite{ML} to
characterize the even-odd ground state energy difference,
\begin{equation}
  \label{eq:ML}
\Delta^{{\rm ML}} (n)= {\cal E}^G_1(n) - 
[ {\cal E}^G_0(n) + {\cal E}^G_0 (n+1)]/2 \, .
\end{equation}
Following the initial g.c.\ studies 
\cite{vD}-
\cite{ML},
the first canonical study  was that of Mastellone, Falci and Fazio (MFF)
\cite{MFF}, who used Lanczos exact diagonalization (with $n \le 12$)
 and a scaling argument to probe the crossover regime.  Berger and
 Halperin (BH) \cite{BH}
showed that essentially the same results could be
 achieved with $n \le 6$ by first reducing the bandwidth and renormalizing
 $\lambda$, thus significantly reducing the calculational effort
involved.  To access larger systems and fully recover the bulk limit,
 fixed-$n$ projected variational BCS wavefunctions (PBCS) were used in
\cite{Braun2} (for $n \le 600$); significant improvements over the
 latter results, in particular in the crossover regime,
 were subsequently achieved in \cite{DS}
 using the density matrix renormalization group (DMRG) (with $n \le 400$).
 Finally, Dukelsky and Schuck \cite{DP} showed that 
a self-consistent RPA approach, that in principle can
be extended to finite temperatures, describes the f.d.\ regime
rather well (though not as well as the DMRG).


To check the quality of the above methods,
we \cite{Braun-thesis,sierra-vd} computed  $E_b^C(n)$ and 
$\Delta^{\rm ML}(n)$ using Richardson's solution 
(Fig.~\ref{fig:exact-gse}). 
 The exact results
(a)
quantitatively agree, for $d \to 0$, with 
the leading $- \Delta^2/2d$
behavior for $E^C_b(n)$ obtained in 
the g.c.\ BCS approach \cite{vD,Braun1,BvD}, which in 
this sense is exact in the bulk limit, corrections being of order $d^0$;
(b) confirm
 that a completely smooth \cite{DS} 
crossover occurs around the scale $d \simeq
 \Delta$ at which the g.c.\ BCS approach breaks down;
(c) show that the PBCS crossover \cite{Braun2}
 is qualitatively correct, but not 
quantitatively, being somewhat too abrupt;
(d) are reproduced remarkably well by the approaches of MFF \cite{MFF} and BH
\cite{BH}; (e) are fully reproduced by the DMRG of \cite{DS} with a relative
error of $< 10^{-4}$ for $n \le 400$; our figures don't show DMRG curves,
since they are indistinghuishable from the exact ones and are discussed in
detail in \cite{DS}.

\section{Conclusions}

The main conclusion we can draw from these comparisons is that the two
approaches based on renormalization group ideas work very well: the
DMRG is essentially exact for this model, but the band-width rescaling
method of BH also gives remarkably (though not quite as) good results
with rather less effort.  In contrast, the PBCS approach is rather
unreliable in the crossover region.

\paragraph{Acknowledgments:\/}
The derivation of the exact solution shown above, which is shorter and
perhaps somewhat more direct than the original published derivation,
was invented by R. W. Richardson too; we thank him for a
private communication suggesting this route. We also thank Moshe
Schechter for helpful comments on the manuscript.


\begin{thebibliography}{99}

\bibitem{RBT}  
D.C.\ Ralph, C.T.\ Black and M.\ Tinkham, Phys.\ Rev.\
  Lett. {\bf 76}, 688 (1996); 78, 4087 (1997).


\bibitem{vD}  
J.\ von Delft {\it et al.}, Phys.\ Rev.\ Lett. {\bf 77}, 3189 (1996).

\bibitem{Braun1}  F.\ Braun {\it et al.}, 
Phys.\ Rev.\ Lett. {\bf 79}, 921 (1997).


\bibitem{BvD}  
F.\ Braun and J.\ von Delft, Phys.\ Rev.\ B {\bf 59}, 9527 (1999).


\bibitem{SA}  
R.A.\ Smith and V.\ Ambegaokar, Phys.\ Rev.\ Lett. {\bf 77}, 4962 (1996).


\bibitem{ML}  
K.A.\ Matveev and A.I.\ Larkin, Phys.\ Rev.\ Lett. {\bf 78}, 3749 (1997).



\bibitem{MFF}  
A.\ Mastellone, G.\ Falci and R.\ Fazio, Phys.\ Rev.\ Lett. {\bf 80},
4542 ( 1998).

\bibitem{BH}  
S.D.\ Berger and B.I.\ Halperin, Phys.\ Rev.\ B {\bf 58}, 5213 (1998).

\bibitem{Braun2}  
F.\ Braun and J.\ von Delft, Phys.\ Rev.\ Lett. {\bf 81}, 4712
(1998).

\bibitem{DS} 
J.\ Dukelsky and G.\ Sierra, Phys.\ Rev.\ Lett. {\bf 83},
  172 (1999); and cond-mat/9906166.

\bibitem{DP}  J.\ Dukelsky and P.\ Schuck, to appear in Phys.\ Lett.\ B.


\bibitem{richardson63a}
R. W.  Richardson, Phys.\ Lett.\ {\bf 3},  277 (1963). 

\bibitem{richardson63b}
R. W.  Richardson, Phys.\ Lett.\ {\bf 5},  82 (1963). 


\bibitem{richardson64}
R. W. Richardson and N. Sherman, Nucl.\ Phys.\ {\bf 52},   221  (1964). 


\bibitem{richardson65a}
R. W. Richardson, 
Phys.\ Lett.\ {\bf 14},   325 (1965). 

\bibitem{richardson65b}
R. W. Richardson,
J.\ Math.\ Phys.\ {\bf 6},   1034  (1965). 

\bibitem{richardson66}
R. W. Richardson,
Phys.\ Rev.\ {\bf 141},   949  (1966).


\bibitem{richardson66-b}
R. W. Richardson,
Phys.\ Rev.\ {\bf 144},   874  (1966).

\bibitem{richardson67}
R. W. Richardson,
Phys.\ Rev.\ {\bf 159}, 792 (1966).

\bibitem{richardson77}
R. W. Richardson, J.\ Math.\ Phys.\ {\bf 18} (1977) 1802.


\bibitem{Braun-thesis}
F.\ Braun, Ph.D.\ thesis, Karlsruhe University (1999);
F. Braun and J.von Delft, 
Advances in Solid State Physics, (Ed. B. Kramer), p.\ 341,
Vieweg, Braunschweig (1999).

\bibitem{sierra-vd}
G. Sierra, J. Dukelsky, G. G. Dussel, J. von Delft, and
F. Braun, cond-mat/9909015.


\bibitem{Soloviev}
V.~G. Soloviev, Mat.\ Fys.\ Skrif.\ Kong.\ Dan.\ Vid.\
  Selsk. {\bf 1}, 1 (1961).


\bibitem{richardson-priv}
R. W. Richardson, private communication (1999).



\end{thebibliography}
\end{document}